\documentclass[reprint,aps,prb,noeprint,notitlepage,nofootinbib,superscriptaddress,longbibliography]{revtex4-2}
\usepackage{amsmath}
\usepackage{amsfonts}
\usepackage{amssymb}
\usepackage{graphicx}
\usepackage{epstopdf}
\usepackage{comment}
\usepackage{physics}
\usepackage{float}
\usepackage{booktabs} 
\usepackage{tabularx}
\usepackage{pifont}
\usepackage{mathrsfs}
\usepackage{makecell}
\usepackage[colorlinks=true]{hyperref}

\begin{document}
\title{Superconducting diode effect in magnetic superconductors realized by nonreciprocal domain-wall dynamics}
\author{Dong Hui Han}
\email{dhhan0726@kaist.ac.kr}
\affiliation{Department of Physics, Korea Advanced Institute of Science and Technology, Daejeon 34141, Republic of Korea}

\author{Suk Bum Chung}
\email{sbchung0@uos.ac.kr}
\affiliation{Department of Physics and Natural Science Research Institute, University of Seoul, Seoul 02504, Republic of Korea}
\affiliation{School of Physics, Korea Institute for Advanced Study, Seoul 02455, Republic of Korea}
\affiliation{Department of Physics and Institute for Condensed Matter Theory, University of Illinois at Urbana-Champaign, Urbana, Illinois 61801, USA
}

\author{Se Kwon Kim}
\email{sekwonkim@kaist.ac.kr}
\affiliation{Department of Physics, Korea Advanced Institute of Science and Technology, Daejeon 34141, Republic of Korea}
\affiliation{Graduate School of Quantum Science and Technology, Korea Advanced Institute of Science and Technology, Daejeon 34141, Republic of Korea}
\affiliation{Department of Physics and Astronomy, Texas A\&M University, College Station, Texas 77843, USA}
\affiliation{Department of Physics, University of Texas at Austin, Austin, Texas 78712, USA}

\begin{abstract}
    A superconducting diode effect is shown to arise in ferromagnetic superconductors through the nonreciprocal dynamics of magnetic domain walls. Specifically, we show that current-driven dynamics of a magnetic domain wall under a certain external field can exhibit a nonreciprocal Walker breakdown, possessing two distinct direction-dependent critical currents beyond which the domain wall precesses continuously. In ferromagnetic superconductors, the constant rotation of a domain wall is shown to give rise to phase slips, opening up dissipation channels, whereby the nonreciprocal Walker breakdown is mapped to the superconducting diode effect. For the nonreciprocal Walker breakdown of a magnetic domain wall, we analytically examine its dependence on the magnetic field and the Gilbert damping and verify the theoretical results with micromagnetic simulations. We then extend the analysis to ferromagnetic superconductors by considering additional effects from the superconductivity and identify criteria for experimental conditions to realize the predicted superconducting diode effect. Our work demonstrates that topological defects, such as domain walls, in magnetic superconductors can serve as an intrinsic nanoscale platform for nonlinear nonreciprocal superconducting functionalities within a single homogeneous material, circumventing the need for complicated engineered heterostructures and thereby enabling the miniaturization of superconducting devices down to the nanometer scale that is challenging to achieve with conventional Josephson junctions.

\end{abstract}

\maketitle

\section{Introduction}

Nonreciprocal transport has long been a central topic in physics, both as a probe of symmetry breaking and for its potential applications in electronic devices such as diodes. Recently, nonreciprocal charge transport in superconductors (SCs) has been intensively explored. A representative example is the superconducting diode effect (SDE), where the critical supercurrent densities beyond which dissipation occurs differ for opposite current directions. This effect is expected to arise from the close interplay between magnetism and superconductivity, which gives rise to magnetoelectric effects~\cite{PhysRevLett.75.2004,doi:10.1126/sciadv.1602390,PhysRevLett.101.107005,szombati2016josephson}. Since diodes constitute fundamental components of electronic circuits, the SDE has attracted significant attention as a potential building block in superconducting electronics. In particular, under cryogenic conditions, conventional semiconducting diodes suffer from several intrinsic limitations, such as the large energy gap, motivating the search for alternative diode mechanisms compatible with superconducting circuits~\cite{neamen2011semiconductor,gui2019review,nadeem2023superconducting,strambini2022superconducting}. To realize the SDE, a system must simultaneously break time-reversal symmetry (TRS) and inversion symmetry (IS)~\cite{shaffer2025theories}. Accordingly, various theoretical approaches have been proposed to achieve the SDE, and the effect has been experimentally observed in several systems~\cite{shaffer2025theories,ando2020observation,PhysRevLett.128.037001,can2021high,Time-reversal,PhysRevB.109.094518,wu2022field,narita2022field,trahms2023diode}.

\begin{figure}[!t]
    \centering
    \includegraphics[width=0.45\textwidth]{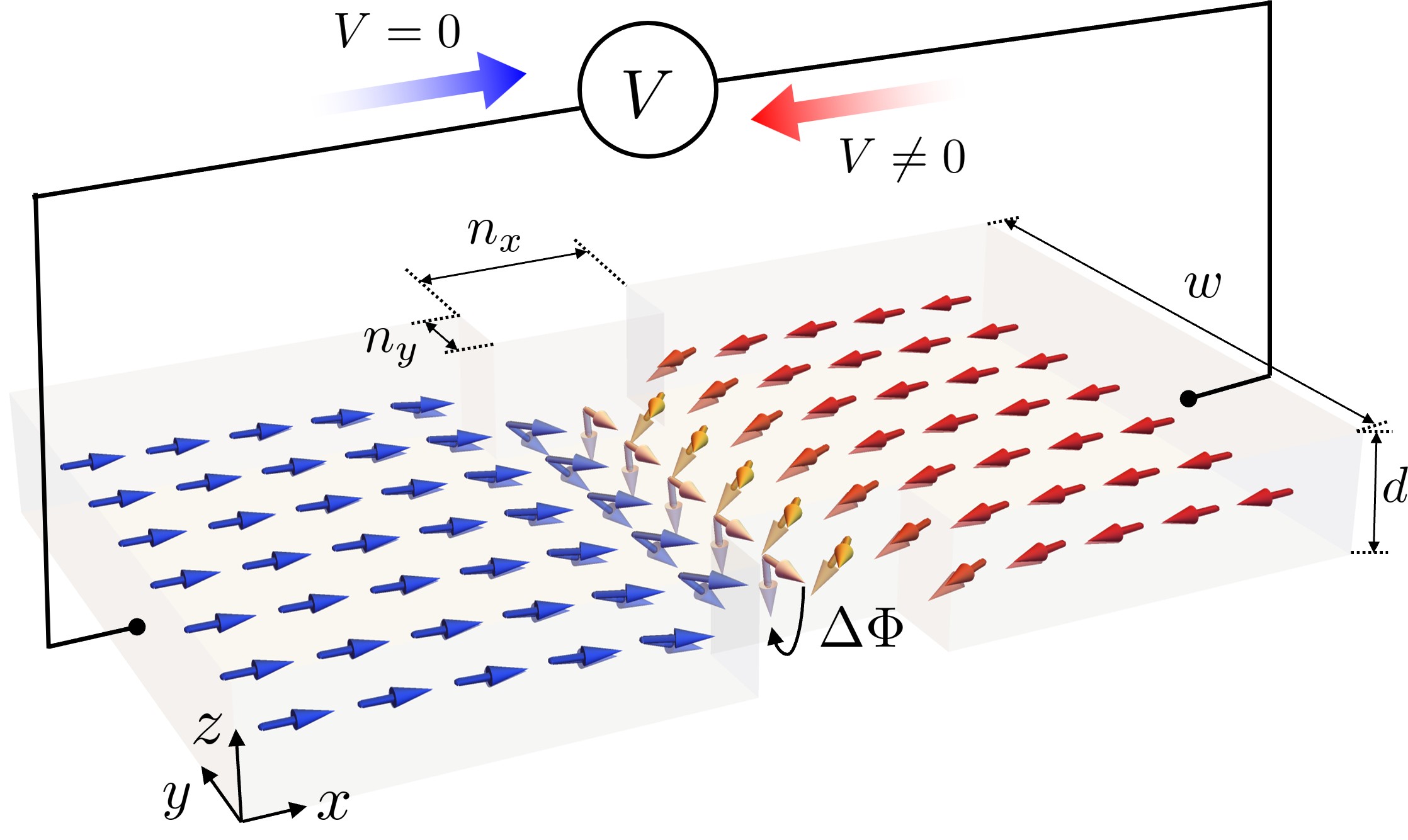} 
    \caption{Schematics of the SDE induced by the DW precession $\Delta\dot{\Phi}$ in the nonreciprocal WB. In the thin film of width $w$ and thickness $d$, the DW is pinned at a notch characterized by the geometric parameters $n_x$ and $n_y$. The measured voltage $V$ originates from the emergent electric fields [Eqs.~(\ref{eq:smf}) and (\ref{eq:emergentfields_electric})]. For the bias current of the equal magnitude, the voltage $V$ across the DW vanishes for one direction (right in the figure) and is finite for the other direction (left in the figure), thereby giving rise to the diode effect.}
    \label{fig1:schematics}
\end{figure}

Here, we propose a method to realize the SDE by exploiting the nonreciprocal dynamics of magnetic textures in SCs. Specifically, we show that the nonreciprocal Walker breakdown (WB) of a supercurrent-driven magnetic domain wall (DW) in ferromagnetic SCs (FMSCs) generates the SDE through phase slips induced by the DW precession. As in other realizations of the SDE, the proposed mechanism requires two ingredients: the dissipation channel and symmetry breaking leading to nonreciprocity. In the present mechanism, the dissipation channel emerges solely through the WB, marking the onset of the DW precession when the driving current density or magnetic field exceeds a critical value. The DW precession is accompanied by an emergent electric field that serves as the primary source of dissipation in SCs where Joule heating is absent and corresponds to the spin-motive force (SMF), the reciprocal counterpart of current-driven magnetization dynamics studied in spintronics~\cite{schryer1974motion,beach2005dynamics,RevModPhys.76.323,jungwirth2016antiferromagnetic,kim2022ferrimagnetic}. Symmetry breaking for nonreciprocity is introduced by external dc magnetic fields, enabling tunable control of the diode effect. Notably, unlike p-n or Josephson junctions, the proposed mechanism requires neither heterostructures nor twist-angle tuning, providing a potentially simpler route to realizing the SDE.

The proposed method for inducing the SDE requires the coexistence of magnetization and superconductivity. Although such coexistence is relatively rare, as magnetization generally suppresses superconductivity by spontaneously breaking TRS, it has been observed in uranium-based heavy-fermion compounds, such as $\mathrm{UGe_2}$, $\mathrm{URhGe}$, and $\mathrm{UCoGe}$~\cite{aoki2019review}. More recently, graphene-based moir\'{e} systems, including twisted bilayer graphene~\cite{doi:10.1126/science.aaw3780,doi:10.1126/science.aav1910}, and rhombohedral trilayer graphene~\cite{zhou2021superconductivity,zhou2021half,mukherjee2025superconducting,han2025signatures} have emerged as promising platforms for such coexistence. These developments motivate phenomenological studies of magnetic textures and their dynamics in FMSCs~\cite{PhysRevB.84.134503,doi:10.1126/sciadv.aat1061,PhysRevLett.122.117002}, to which the present work contributes. In this work, we suggest an approach to induce the SDE via the nonreciprocal WB of magnetic DWs in FMSCs. Given the equivalence between the equations of motion (EOMs) of magnetization in ferromagnets (FMs) and in FMSCs under the incompressible limit, we first investigate the nonreciprocal WB of the current-driven DW in FMs and derive the resulting nonreciprocal current-voltage (I-V) characteristics from the SMF. We then demonstrate the equivalence of the EOMs using a hydrodynamic formalism for FMSCs, show that the SDE can emerge through phase slips from the DW precession analogous to the SMF in FMs, and identify the conditions required to extend the results obtained in FMs to FMSCs.

In Sec.~\ref{sec:WBinFM}, we present a theory of nonreciprocal WB of ferromagnetic DWs for both unpinned and notch-pinned cases and verify the theory by comparison with micromagnetic simulations. In Sec.~\ref{sec:SDEinFMSCs}, we demonstrate the emergence of the SDE by identifying the equivalence between the EOMs in FMs and FMSCs within the $\mathbf{d}$-vector formalism, and outline several conditions necessary for the extension of the results from FMs to FMSCs. We summarize and discuss our results in Sec.~\ref{sec:conclusion}.

\section{Nonreciprocal Walker breakdown in ferromagnets}
\label{sec:WBinFM}

\subsection{Theory}

We consider a quasi-one-dimensional DW in a ferromagnetic thin film. The Hamiltonian density is given by
\begin{equation}
    \mathcal{H}=\frac{A}{2}(\partial_x \mathbf{m})^2-\frac{K}{2}m_x^2+\frac{K_1}{2}m_z^2-M\mathbf{H}\cdot\mathbf{m},\label{eq:Hamil_fm}
\end{equation}
where $A$ is a ferromagnetic exchange constant, $K$ is an easy-axis anisotropy constant, $K_1$ is a hard-axis anisotropy constant, $M$ is a magnetization, $\mathbf{H}$ is an external magnetic field, and the unit magnetization vector $\mathbf{m}$ is given by
\begin{equation}
    \mathbf{m}=(m_x,m_y,m_z)=(\cos{\theta},\sin{\theta}\cos{\phi},\sin{\theta}\sin{\phi})
\end{equation}
with polar and azimuthal angles, $\theta$ and $\phi$. The dynamics of magnetization in FMs obeys the Landau-Lifshitz-Gilbert~(LLG) equation, 
\begin{equation}
    \mathcal{J}(\dot{\mathbf{m}}-\alpha\mathbf{m}\times\dot{\mathbf{m}})=-\frac{\delta \mathcal{H}}{\delta\mathbf{m}}\times\mathbf{m}+\mathcal{P}(1-\beta\mathbf{m}\times)(\mathbf{j}\cdot\nabla)\mathbf{m},\label{eq:llg}
\end{equation}
where the last term represents the adiabatic and nonadiabatic spin-transfer torque (STT)~\cite{tserkovnyak2008theory}. Here, $\alpha$ is the damping parameter, $\beta$ characterizes the nonadiabatic STT, $\mathcal{J}=M/\gamma$ is the spin density, $\gamma$ is the gyromagnetic ratio, $\mathbf{j}=j\hat{\mathbf{x}}$ denotes the current density, and $\mathcal{P}=\hbar(\sigma_\uparrow-\sigma_\downarrow)/2e(\sigma_\uparrow+\sigma_\downarrow)$ is the spin-conversion factor with the electric conductivity of spin-up (down) electrons $\sigma_\uparrow\,(\sigma_\downarrow)$ and the charge of electron $e$~\cite{tserkovnyak2008theory}.

Exploiting the Walker ansatz~\cite{schryer1974motion,kim2023mechanics},
\begin{equation}
    \tan{\frac{\theta(x,t)}{2}}=\exp\left[\frac{x-X(t)}{\lambda}\right],\quad \phi(x,t)=\Phi(t),
\end{equation}
the low-energy dynamics of the DW can be described with two collective coordinates, $X$ and $\Phi$, corresponding to the DW position and the azimuthal angle measured in the $y$-$z$ plane, respectively. Here $\lambda=\sqrt{A/K}$ denotes the DW width. 
From the LLG equation [Eq.~(\ref{eq:llg})], the EOMs of the collective coordinates are derived as~\cite{A.Thiaville_2005}
\begin{align}
\left(\alpha+\frac{1}{\alpha}\right)\dot{\Phi}&=-\frac{1}{2\mathcal{J}}[K_1\sin{2\Phi}-\pi M( H_z \cos{\Phi}\label{eq:fm_EOM1_nonotch}\\
&\quad-H_y \sin{\Phi})]+\frac{\mathcal{P} j}{\mathcal{J}\lambda}\left(1-\frac{\beta}{\alpha}\right),\nonumber\\
-\frac{\dot{X}}{\lambda}&=\frac{\dot{\Phi}}{\alpha}+\frac{\beta\mathcal{P}j}{\mathcal{J}\alpha\lambda}.\label{eq:fm_EOM2_nonotch}
\end{align}

\begin{figure}[t]
    \centering
    \includegraphics[width=0.9\linewidth]{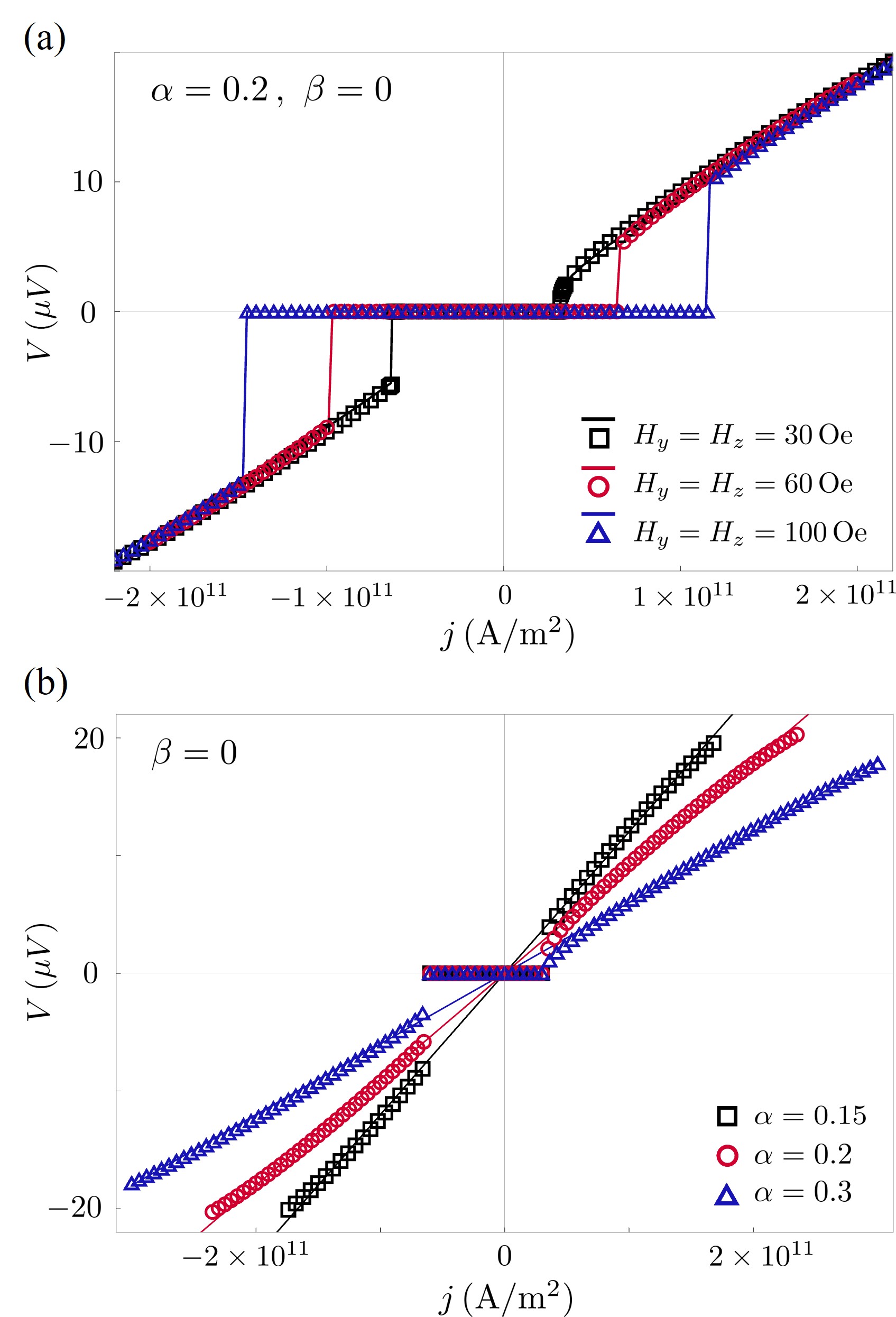} 
    \caption{I-V curves of a pinned DW in a notch obtained from simulations with $H_y=H_z$ and $\beta=0$. (a) shows the dependence on the magnetic field intensity where $\alpha=0.2$. The solid lines represent the numerical solutions from the EOMs [Eqs.~(\ref{eq:fm_EOM1})~and (\ref{eq:fm_EOM2})], which are used to determine $k_p=0.45\times 10^{-3}\,\mathrm{N/m}$. (b) shows the dependence on $\alpha$ where $H_y=H_z=30\,\mathrm{Oe}$. The solid lines represent $\langle V\rangle=-2\mathcal{P}\langle\dot{\Phi}\rangle$ [Eqs.~(\ref{eq:smf}) and (\ref{eq:avgs})]. For $\alpha=0.15$ and $\alpha=0.2$, the DW is depinned from the notch at $j_{\mathrm{dep}}=1.74\times 10^{11}\mathrm{A/m^2}$ and $j_{\mathrm{dep}}=2.35\times 10^{11}\mathrm{A/m^2}$, respectively.}
    \label{fig:ivcurve}
\end{figure}

To obtain an analytical expression for the diode efficiency $\eta$, defined as 
\begin{equation}
    \eta=\frac{\left|j_c^+\right|-\left|j_c^-\right|}{\left|j_c^+\right|+\left|j_c^-\right|}\label{eq:efficiency},
\end{equation}
we focus on the case $H_y=H_z>0$, and neglect the nonadiabatic effects, i.e., $\beta=0$, which is addressed in Appendix~\ref{app:nonadiabatic}. Here, $j_c^+$ and $j_c^-$ denote the critical current densities for the WB when the current flows in the positive and the negative $x$ directions, respectively. When the bias current exceeds $j_c^\pm$, the DW enters the WB regime, where the DW continuously precesses $\dot{\Phi}\neq0$.
Since the EOM of $\Phi$ is in the overdamped regime~\cite{0_fmsc_footnote}, the critical current densities can be obtained by analyzing the stationary solutions of the EOMs~[Eqs. (\ref{eq:fm_EOM1_nonotch}) and (\ref{eq:fm_EOM2_nonotch})] under the condition $\dot{\Phi}=\dot{X}=0$, yielding
\begin{align}
    j_c^+&=
    \begin{cases}
        \frac{\lambda}{2\mathcal{P}}\left(K_1+\mathcal{E}_Z^2/4K_1\right)\,& (H_z<H_c)\\
        \frac{\lambda}{2\mathcal{P}}\left(-K_1+\sqrt{2}\mathcal{E}_Z\right), \,& (H_z>H_c)\label{eq:jp}
    \end{cases}\\
    j_c^-&=-\frac{\lambda}{2\mathcal{P}}\left(K_1+\sqrt{2}\mathcal{E}_Z\right).\label{eq:jm}
\end{align}
The results are analogous to the critical current densities obtained from the phase dynamics in a Josephson junction without capacitance~\cite{PhysRevB.109.094518}.
Here, $\mathcal{E}_Z=\pi M H_z$ is proportional to the Zeeman energy density, and $H_c=2\sqrt{2}K_1/\pi M$ represents the characteristic field at which the dominant contribution to the precessional motion of the DW crosses over from $K_1$ to $\mathbf{H}$. Note that the subdominant second term of $j_c^+$ becomes independent of $K_1$ for $H_z>H_c$. From these results, it is straightforward to obtain the analytical expressions for the diode efficiency $\eta$,
\begin{figure*}[t]
    \centering
    \includegraphics[width=1\linewidth]{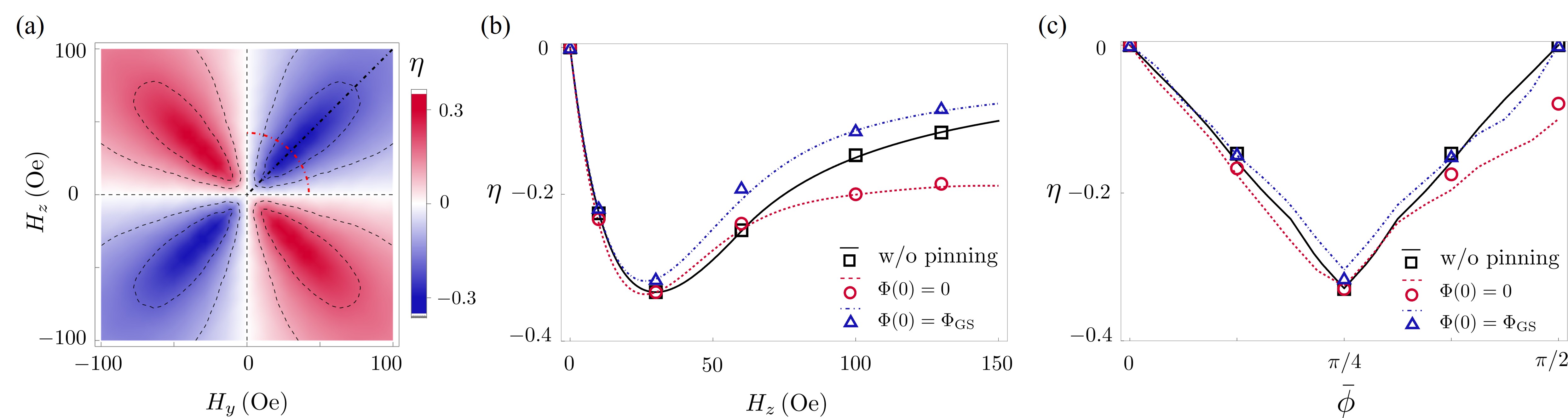} 
    \caption{Plots of the diode efficiency $\eta$ [Eq.~(\ref{eq:efficiency})] for various magnetic field conditions. (a) Numerical results obtained from the EOMs [Eqs.~(\ref{eq:fm_EOM1_nonotch})~and~(\ref{eq:fm_EOM2_nonotch})] for $\abs{H_y},\,\abs{H_z}\leq 100\,\mathrm{Oe}$, where $\alpha=0.2$ and $\beta=0$. The black dashed lines indicate contours of $\eta=0,\,0.1,\,0.2$. (b) and (c) present comparisons with micromagnetic simulations for the black dash-dotted line $H_y=H_z>0$ and the red dash-dotted line $(H_y,H_z)=\sqrt{2}H_{\mathrm{max}}(\cos{\bar{\phi}},\sin{\bar{\phi}})$ in (a), respectively. Here $\bar{\phi}$ denotes the azimuthal angle in the $y$-$z$ plane. In both panels, the solid, dotted, and dash-dotted lines represent numerical solutions of the EOMs without pinning, with pinning and the initial condition $\Phi(t=0)=0$, and with pinning and initialized in the ground state, i.e. $\Phi(0)=\Phi_{\mathrm{GS}}$, respectively. Simulation results are shown as the square, circular, and triangular symbols. Here $\Phi_{\mathrm{GS}}$ denotes the value of $\Phi$ that minimizes the magnetostatic energy under magnetic fields, and the solid line in (b) coincides with the exact analytical results given in Eq.~(\ref{eq:analytic_effi}).}
    \label{fig:ratio}
\end{figure*}

\begin{equation}
    \eta=
    \begin{cases}
        \frac{\mathcal{E}_Z(\mathcal{E}_Z-4\sqrt{2}K_1)}{(\mathcal{E}_Z+2\sqrt{2}K_1)^2}\,& (H_z<H_c)\\
        -\frac{K_1}{\sqrt{2}\mathcal{E}_Z}\,& (H_z>H_c),\label{eq:analytic_effi}
    \end{cases}
\end{equation}
and the magnetic field intensity that maximizes $\eta$ is given by 
\begin{equation}
    H_{\mathrm{max}}=\frac{\sqrt{2}K_1}{\pi M}=\frac{H_c}{2},\label{eq:max_effi_H}
\end{equation} 
at which $\eta$ attains its maximum value,
\begin{equation}
    \eta_{\mathrm{max}}=\eta(H_y=H_z=H_{\mathrm{max}})=\frac{1}{3}.
\end{equation}
Combined with the adiabatic SMF generated by the precessional motion of the DW, written as~\cite{PhysRevB.77.134407}
\begin{equation}
    V=-2\mathcal{P}\dot{\Phi},\label{eq:smf}
\end{equation}  
the nonreciprocal WB results in asymmetric I-V characteristics, giving rise to a diode effect.
This constitutes our first main result: analytical expressions for the diode efficiency $\eta$ and its maximum value, and a theoretical demonstration of the diode effect in the adiabatic SMF for the case $H_y=H_z>0$.

At this point, we provide a few remarks. First, Eqs. (\ref{eq:jp}) and (\ref{eq:jm}) become reciprocal for $K_1=0$ or $H_y=H_z=0$, and $\eta_{\mathrm{max}}$ is given by $1/3$, independent of material parameters. This behavior is equivalent to the thermodynamic diode effect in twisted nodal SCs, where TRS is explicitly broken by external magnetic fields and the second-harmonic of Josephson effect is associated with $K_1$~\cite{PhysRevB.109.094518}. 
Secondly, to extend the present analysis to the case $H_y\neq H_z$, we consider two extreme cases in which one of the magnetic fields vanishes. Guided by the analogy with twisted nodal SCs, we find that a finite $\eta$ requires the simultaneous application of two magnetic fields to break the symmetries. Specifically, in the absence of the magnetic fields, the present system possesses two degenerate ground states, analogous to those of twisted nodal SCs with the twist angle close to $45^\circ$. The difference is that, unlike the latter, whose degenerate free-energy minima are located at finite phase differences, one of the ground states in the present work is located at the origin, $\Phi=0$. To induce the diode effect, both TRS and mirror symmetries must be broken. The symmetries act on the unit magnetization as 
\begin{align}
    \mathcal{T}\,\,&:\,\,(m_x,m_y,m_z)\rightarrow(-m_x,-m_y,-m_z),\\
    \mathcal{M}_x&:\,\,(m_x,m_y,m_z)\rightarrow(m_x,-m_y,-m_z).
\end{align}
The two magnetic fields break TRS, but they play distinct additional roles. Since $\Phi$ denotes the azimuthal angle of the DW on the $y$-$z$ plane, the mirror operations associated with the diode effect arising from the DW precession are $\mathcal{M}_y$ and $\mathcal{M}_z$, which are broken by $H_z$ and $H_y$, respectively. In the present case, the hard-axis anisotropy along the $z$-axis stabilizes the N\'{e}el-type DW whose center magnetization lies along the $y$-axis, making $\mathcal{M}_y$ the mirror symmetry responsible for the nonreciprocity~\cite{2_fmsc_footnote}. Consequently, $H_z$ breaks the mirror symmetry while preserving the ground-state degeneracy, whereas $H_y$ lifts the degeneracy as explicitly breaking TRS. These observations suggest that, as in the thermodynamic diode effect demanding the explicitly broken TRS, both magnetic fields are required to induce the diode effect and the condition $H_y=H_z$ is optimal for maximizing $\eta$.

Lastly, we incorporate pinning effects into our analysis, which are commonly realized in experiments through various mechanisms, such as defects and voltage-controlled magnetic anisotropy~\cite{PhysRevB.98.054406,tan2019high}. We consider a DW pinned by a notch geometry~\cite{PhysRevB.83.174444}, which suppresses its translational motion and enables it to function as an immobile diode. The effective magnetic field arising from the pinning potential $V_p=k_p X^2/2$, approximated by a quadratic function, is given by 
\begin{equation}
    H_p(X)=-\frac{1}{2M w d}\frac{\partial V_p}{\partial X} =-\frac{k_p X}{2M w d},\label{eq:pinningpotential}
\end{equation}
where $w$ and $d$ denote the width and thickness, and the pinning strength $k_p$ depends on the shape of notch in Fig.~(\ref{fig1:schematics}) and is determined through a fitting procedure~\cite{PhysRevB.83.174444,voto2017pinned} [see Fig.~\ref{fig:ivcurve}(a)]. The EOMs of the collective coordinates, augmented by the pinning potential, are written as

\begin{align}
    \left(\alpha+\frac{1}{\alpha}\right)\dot{\Phi}&=-\frac{1}{2\mathcal{J}}[K_1\sin{2\Phi}-\pi M( H_z \cos{\Phi}\label{eq:fm_EOM1}\\
    &\quad-H_y \sin{\Phi})]+\frac{\gamma H_p}{\alpha}+\frac{\mathcal{P} j}{\mathcal{J}\lambda}\left(1-\frac{\beta}{\alpha}\right),\nonumber\\
    -\frac{\dot{X}}{\lambda}&=\frac{\dot{\Phi}}{\alpha}+\frac{\beta\mathcal{P}j}{\mathcal{J}\alpha\lambda}-\frac{\gamma H_p}{\alpha}.\label{eq:fm_EOM2}
\end{align}

Here, we point out a possible effect of pinning on the dynamics. In the absence of pinning, the dynamics of $\Phi$, governed by the EOMs [Eqs. (\ref{eq:fm_EOM1_nonotch}) and (\ref{eq:fm_EOM2_nonotch})], lies in the overdamped regime, where inertia vanishes, and the local slope of the effective potential immediately determines $\dot{\Phi}$. By contrast, the introduction of pinning generates an effective inertia, thereby rendering the dynamics dependent on the initial conditions and enabling an additional mechanism leading to a finite $\eta$, namely a dynamical diode effect~\cite{PhysRevB.109.094518}. In the next subsection, we compare the theoretical results and predictions with micromagnetic simulations and further elucidate the underlying physics.

\subsection{Simulation results and discussion}

We perform micromagnetic simulations using MuMax3, an open-source software solving the LLG equation~\cite{vansteenkiste2014design}. The material parameters used in the simulations are $A=20\,\mathrm{pJ/m}$, $K=0.4\,\mathrm{MJ/m^3}$, $K_1=2\,\mathrm{kJ/m^3}$, $M=0.3\,\mathrm{MA/m}$, $\abs{\gamma}=1.76\times10^{11}\,\mathrm{rad/s\,T}$, $\lambda=7\,\mathrm{nm}$, $\mathcal{P}=\hbar/2e$, and $\alpha=0.2$~\cite{PhysRevB.83.174444}. The geometry of the system is specified by $w=\mathrm{60\,nm}$, $d=\mathrm{1\,nm}$, $n_x=\mathrm{15\,nm}$, and $n_y=\mathrm{7.5\,nm}$. In simulations and numerical calculations of the EOMs, unless otherwise noted, the system is initialized in its ground state under the magnetic fields prior to the application of the bias current, and, for these simulation parameters, $H_{\mathrm{max}}=30\,\mathrm{Oe}$. 

Figure \ref{fig:ivcurve}(a) displays I-V characteristics of a pinned DW under different magnetic field strengths. As predicted by Eqs.~(\ref{eq:jp}) and (\ref{eq:jm}), $\abs{j_c^\pm}$ increases with the magnetic fields. By contrast, the I-V curves converge to a single straight line in the WB regime, regardless of the field strength. This behavior can be understood as follows. A DW trapped in a notch is displaced from the center and oscillates with small amplitudes driven by the STT~\cite{PhysRevB.83.174444,voto2017pinned}. As a result, Eqs.~(\ref{eq:fm_EOM1}) and (\ref{eq:fm_EOM2}) yield the time-averaged displacement from the center $\expval{X}$ and the angular frequency $\langle{\dot{\Phi}}\rangle$~\cite{1_fmsc_footnote}:
\begin{equation}
    \expval{X}=-\frac{2w d \mathcal{P} j}{k_p \alpha\lambda},\quad\quad\langle{\dot{\Phi}}\rangle=\frac{\mathcal{P}j}{\mathcal{J}\alpha \lambda}.\label{eq:avgs}
\end{equation}
Here $\expval{\dots}$ denotes the time average. The latter expression is linear in $j$ and independent of the magnetic field strength, in agreement with the simulation results. Note that the first equation allows us to roughly estimate the depinning current density, $j_{\mathrm{dep}}$, by assuming that the depinning occurs when $\langle X\rangle$ exceeds $n_x/2$. This criterion yields 
\begin{equation}
    j_{\mathrm{dep}}\sim \frac{n_x k_p \alpha \lambda}{4wd\mathcal{P}}.\label{eq:depin}
\end{equation}
For $\alpha=0.2$, this gives $j_{\mathrm{dep}}=1.21\times 10^{11}\,\mathrm{A/m^2}$, which is smaller than $j_{\mathrm{dep}}$ observed in simulations [see Fig.~\ref{fig:ivcurve}(b)], but captures its order of magnitude correctly. Figure~\ref{fig:ivcurve}(b) shows the $\alpha$ dependence of the I-V characteristics. In contrast to Eq.~(\ref{eq:avgs}), the time-averaged angular frequency without pinning $\langle{\dot{\Phi}}\rangle_0$ is given by 
\begin{equation}
    \langle{\dot{\Phi}}\rangle_0=\frac{\mathcal{P}\alpha j}{\mathcal{J}\lambda(1+\alpha^2)},\label{eq:avg_w/o_pinning}
\end{equation}
which is approximately proportional to $\alpha$ when the damping is sufficiently small $\alpha \ll 1$. This discrepancy originates from the difference in the dominant mechanism governing the DW precession. The pinning potential gives rise to $\mathbf{H}_p$, which can be interpreted as a position-dependent magnetic field along the $x$-axis, i.e., $\mathbf{H}_p\propto X\hat{x}$. When the DW is displaced to a position $\expval{X}\propto j/\alpha$ by the STT, this effective field becomes the primary source determining the precession frequency. By contrast, in the absence of pinning, the precession is dominated by the damping torque, which is proportional to $\alpha$, resulting in a reduced $V$ by a factor of $\alpha^2$ compared with the pinned case. These analyses are in good agreement with the simulation results, which show that the slopes of I-V curves scale inversely with $\alpha$.

Figure~\ref{fig:ratio}(a) presents the magnetic field dependence of $\eta$ obtained by numerically solving the EOMs [Eqs.~(\ref{eq:fm_EOM1_nonotch}) and (\ref{eq:fm_EOM2_nonotch})] without pinning effects. By symmetry, reversing the direction of magnetic field changes the sign of $\eta$, and the absolute value of $\eta$ is maximized for $\abs{H_y}=\abs{H_z}$ at fixed field magnitude $\abs{\mathbf{H}}$, along which $\abs{\eta}$ attains its global maximum. The simulation results shown in Figs.~\ref{fig:ratio}(b) and \ref{fig:ratio}(c) are in good agreement with these findings. Furthermore, Fig.~\ref{fig:ratio}(c) shows that a finite $\eta$ emerges for $\Phi(t=0)=0$ even when only $H_z$ is applied. This result demonstrates that the initial conditions of the pinned DWs have a substantial impact on $\eta$, providing direct evidence for the existence of inertia in the pinned DW. Such inertia can give rise to a dynamical diode effect, previously discussed in the context of the Josephson diode effect~\cite{PhysRevB.109.094518}. From a symmetry perspective, the application of $H_z$ is sufficient to allow the dynamical diode effect. However, as shown in Fig.~\ref{fig:ratio}(c), the resulting effect remains negligible since the system is still in an almost overdamped regime, i.e., weak inertia. One way to enhance the inertia effect, including the dynamical diode effect, is to reduce $\alpha$; however, this also makes it more difficult to trap the DW in the notch geometry~[see Eq.~(\ref{eq:depin})]. These results suggest a possible route to control $\eta$ by tuning the magnetic field conditions and the notch geometry, which may be used for future applications. This is our second main result: the verification of the theory through comparison with micromagnetic simulations and the analysis of the dynamics of the pinned DW in the nonreciprocal WB, as shown in Fig.~\ref{fig:ivcurve} and Fig.~\ref{fig:ratio}. In the next section, we extend the present analysis to FMSCs.

\section{Superconducting diode effect in ferromagnetic superconductors}
\label{sec:SDEinFMSCs}
\subsection{Hydrodynamic theory of FMSCs}
We begin by briefly reviewing the hydrodynamics of FMSCs developed in previous works~\cite{PhysRevResearch.3.013051,PhysRevB.108.134509,dao2025,RevModPhys.47.331}. This framework shows that, under the incompressible limit, the supercurrent-driven magnetization dynamics in FMSCs is equivalent to that in FMs, allowing the SDE to arise from the nonreciprocal WB as in FMs. The starting point is the superconducting order parameter. In general, the gap functions $\Delta_{s,s'}$, the order parameters in SCs, are represented within $\mathbf{d}$-vector formalism as $\Delta_{s,s'}\equiv i(\mathbf{d}\cdot\boldsymbol{\sigma}\sigma^y)_{s,s'}$, where $s$ and $s'$ represent the spin degrees of freedom, and $\{\sigma^i\}$ denote the Pauli matrices.
By introducing the $\mathbf{d}$-vector,
\begin{equation}
    \mathbf{d}=\sqrt{\frac{\rho}{2}}\frac{e^{i\phi}(\hat{\mathbf{u}}+i\hat{\mathbf{v}})}{\sqrt{2}},\label{eq:orderparmeter}
\end{equation}
we can express several physical quantities in a convenient form, e.g., the Cooper pair number density $\rho=2\mathbf{d}\cdot\mathbf{d}^*$ and the spin polarization $\mathbf{s}=2i\mathbf{d}\times\mathbf{d}^*$. Here, $\phi$ denotes the superconducting phase. In this work, we focus on fully spin-polarized triplet SCs, for which $\hat{\mathbf{u}}$ and $\hat{\mathbf{v}}$ are mutually perpendicular unit vectors and $\mathbf{d}\cdot \mathbf{d}=0$ is satisfied, thereby maximizing $\mathbf{s}$. Given the order parameter, a phenomenological Ginzburg-Landau free energy density, expanded up to second order in derivatives, can be written as~\cite{PhysRevResearch.3.013051,PhysRevB.108.134509}
\begin{align}
    \mathcal{F}=&\frac{A_c \rho}{2\rho_0}\left\{\left[\partial_i\phi-\frac{q}{\hbar}A_i-\hat{\mathbf{s}}\cdot(\hat{\mathbf{u}}\times\partial_i \hat{\mathbf{u}})\right]^2+\frac{(\partial_i\hat{\mathbf{s}})^2}{2}\right\}\\
    &+\rho\left[qV_e-\gamma\hbar\mathbf{\hat{s}}\cdot\mathbf{H}\right]+\frac{\rho^2}{\rho_0^2}\mathcal{F}_m+\mathcal{F}_\rho,\nonumber
\end{align}
where the subset of magnetic free energy density contributions $\mathcal{F}_m$ is given by 
\begin{equation}
    \mathcal{F}_m=\frac{A_s}{2}\abs{\nabla \hat{\mathbf{s}}}^2-\frac{\bar{K}}{2}(\hat{\mathbf{s}}\cdot\hat{x})^2+\frac{\bar{K}_1}{2}(\hat{\mathbf{s}}\cdot\hat{z})^2.
\end{equation}
Here, $A_s$ denotes the excess spin stiffness, $\bar{K}$ ($\bar{K}_1$) plays the same role as $K$ ($K_1$) in Eq.~(\ref{eq:Hamil_fm}), $\mathcal{F}_\rho$ represents the contribution from fluctuations of $\rho$, $q$ is the charge of Cooper pair, $A_c$ denotes the charge stiffness, $A_i$ represents the vector potential component, $V_e$ is the scalar potential, $\rho_0$ denotes the Cooper pair density in the absence of fluctuations, and $\hat{\mathbf{s}}=\hat{\mathbf{u}}\times\hat{\mathbf{v}}$ is a unit vector of $\mathbf{s}$. Assuming the incompressible (London) limit~\cite{PhysRevResearch.3.013051,PhysRevB.108.134509}, i.e., $\rho=\rho_0$, the free energy density can then be recast as
\begin{align}
    \mathcal{\tilde{F}}=& \frac{A_c}{2}\left[\partial_i\phi-\frac{q}{\hbar}A_i-\hat{\mathbf{s}}\cdot(\hat{\mathbf{u}}\times\partial_i \hat{\mathbf{u}})\right]^2+\mathcal{\tilde{F}}_m+q\rho_0V_e\label{eq:freeenergy}
\end{align}
with the magnetic free energy density $\mathcal{\tilde{F}}_m$:
\begin{equation}
    \mathcal{\tilde{F}}_m=\frac{\bar{A}}{2}\abs{\nabla \hat{\mathbf{s}}}^2-\frac{\bar{K}}{2}(\hat{\mathbf{s}}\cdot\hat{x})^2+\frac{\bar{K}_1}{2}(\hat{\mathbf{s}}\cdot\hat{z})^2-\gamma\hbar\rho_0\mathbf{\hat{s}}\cdot\mathbf{H},
\end{equation}
where $\bar{A}=A_s+A_c/2$ is the total spin stiffness, which is equivalent to $A$ in Eq.~(\ref{eq:Hamil_fm}). The charge supercurrent density $\mathbf{j}_s=(j_{s,x},j_{s,y},j_{s,z})$ is then given by 
\begin{equation}
    j_{s,i}=-\frac{\delta \mathcal{\tilde{F}}}{\delta A_i}=\frac{q A_c }{\hbar}\left[\partial_i\phi-\frac{q}{\hbar}A_i-\hat{\mathbf{s}}\cdot(\hat{\mathbf{u}}\times\partial_i \hat{\mathbf{u}})\right]. \label{eq:supercurrent}
\end{equation}
Note that $\rho$ drops out of the expression due to the cancellation in the incompressible limit.

To describe the supercurrent-driven DW dynamics in FMSCs and establish its analogy to that in FMs [Eq.~(\ref{eq:llg})], we employ a Lagrangian formalism to derive the EOM of the magnetization. Starting from the Lagrangian density with the kinetic term $\mathcal{K}$~\cite{PhysRevResearch.3.013051,PhysRevB.108.134509,dao2025},
\begin{equation}
    \mathcal{L}=\mathcal{K}-\mathcal{F},\label{eq:lagrangian}
\end{equation}
we can derive the hydrodynamic EOMs for the three variables, $\rho$, $\phi$, and $\hat{\mathbf{s}}$, with the details of the derivations given in Ref.~\cite{PhysRevB.108.134509}. These EOMs represent the charge conservation, the Josephson relation, and the dynamics of the spin direction, respectively. By incorporating dissipation~\cite{1353448}, the last equation can be recast as
\begin{equation}
    \hbar\rho_0\left(\partial_t\hat{\mathbf{s}}+\frac{\mathbf{j}_s\cdot\nabla}{q\rho_0}\hat{\mathbf{s}}+\alpha\hat{\mathbf{s}}\times\partial_t \hat{\mathbf{s}}\right)=-\hat{\mathbf{s}}\times \frac{\delta \tilde{\mathcal{F}}_m}{\delta \hat{\mathbf{s}}},\label{eq:spinllg}
\end{equation}
which becomes equivalent to the LLG equation [Eq.~(\ref{eq:llg})] with the maximum $\mathcal{P}$ and $\beta=0$ under the identifications: $\hbar\rho_0\mapsto \mathcal{J}$, $q\mapsto -2e$, $\mathbf{j}_s\mapsto \mathbf{j}$, and $\mathbf{\hat{s}}\mapsto-\mathbf{m}$. This implies that the DW dynamics in FMSCs can be straightforwardly obtained from that in FMs discussed in the previous section and, as in FMs, nonreciprocal WB emerges in the presence of finite hard-axis anisotropy and external dc magnetic fields, leading to nonzero $j_c^\pm$ and nonreciprocity~[Eqs.~(\ref{eq:jp}) and (\ref{eq:jm})].

This nonreciprocal WB in FMSCs, where dissipation\textemdash particularly Joule heating in normal metals\textemdash is absent, gives rise to the SDE when the DW precession generates a voltage corresponding to the SMF in FMs~[Eq.~(\ref{eq:smf})]. In FMSCs, this voltage originates from emergent gauge fields associated with $\mathrm{U(1)_{\phi+s}}$ order parameter redundancy~\cite{G_E_Volovik_1987,vollhardt2013superfluid,PhysRevB.108.134509}. This redundancy reflects the fact that the order parameter $\mathbf{d}$ [Eq.~\ref{eq:orderparmeter}] remains invariant under the two simultaneous operations: a rotation of the unit vectors, $\hat{\mathbf{u}}$ and $\hat{\mathbf{v}}$, about $\hat{\mathbf{s}}$, and a shift of $\phi$, analogous to the gauge transformation in classical electromagnetism~\cite{1998clel}. The associated gauge fields give rise to emergent electromagnetic fields, which constitute the primary dissipation mechanism in the present work. Their explicit forms are derived as follows. The kinetic term $\mathcal{K}$ in Eq.~(\ref{eq:lagrangian}), which satisfies the commutation relation $[d^*_i(\mathbf{r}),d_j(\mathbf{r}')]=2i\hbar \delta(\mathbf{r}-\mathbf{r}')$, is given by~\cite{PhysRevB.108.134509} 
\begin{equation}
    \mathcal{K}=2i\hbar\mathbf{d}^*\cdot\partial_t\mathbf{d}=-\hbar\rho[\partial_t\phi-\hat{\mathbf{s}}\cdot(\hat{\mathbf{u}}\times\partial_t \hat{\mathbf{u}})].\label{eq:kineticenergy}
\end{equation}
In the incompressible limit, using the Eqs.~(\ref{eq:freeenergy}) and (\ref{eq:kineticenergy}), the Lagrangian density can be recast as
\begin{align}
    \mathcal{L}=&-\hbar\rho_0\left[\partial_t\phi+\frac{q}{\hbar}V_e-\hat{\mathbf{s}}\cdot(\hat{\mathbf{u}}\times\partial_t \hat{\mathbf{u}})\right]\nonumber \\
    &-\frac{A_c}{2}\left[\partial_i\phi-\frac{q}{\hbar}A_i-\hat{\mathbf{s}}\cdot(\hat{\mathbf{u}}\times\partial_i \hat{\mathbf{u}})\right]^2-\mathcal{\tilde{F}}_m\label{eq:Lagrangian_rev},
\end{align}
which allows us to identify the emergent gauge fields, $\mathcal{V}_e$ and $\mathcal{A}_i$:
\begin{align}
    \mathcal{V}_e&=-\frac{\hbar}{q}\hat{\mathbf{s}}\cdot(\hat{\mathbf{u}}\times\partial_t \hat{\mathbf{u}}),\\
    \mathcal{A}_i&=\frac{\hbar}{q}\hat{\mathbf{s}}\cdot(\hat{\mathbf{u}}\times\partial_i \hat{\mathbf{u}}).
\end{align}
These emergent gauge fields then give rise to emergent electromagnetic fields, $e_i$ and $b_i$:
\begin{align}
    e_i&=-\frac{\hbar}{q}\hat{\mathbf{s}}\cdot (\partial_i \hat{\mathbf{s}}\times \partial_t \hat{\mathbf{s}}),\label{eq:emergentfields_electric}\\
    b_i&=-\frac{\hbar\epsilon_{ijk}}{2q} \hat{\mathbf{s}}\cdot (\partial_j \hat{\mathbf{s}}\times \partial_k \hat{\mathbf{s}}).\label{eq:emergentfields_magnetic}
\end{align}
This emergent electric field has an identical form to the adiabatic SMF~\cite{PhysRevB.77.134407}, resulting in the same voltage [Eq.~\ref{eq:smf}]. Note that this voltage can be interpreted as a constant rate of $4\pi$ phase slips, provided that rotations around $\hat{\mathbf{s}}$ are equivalent to the shift of $\phi$~\cite{PhysRevB.108.134509}.

\subsection{Criteria for extending the theory in FMs}

In this subsection, we clarify the regime in which the extension is valid and identify the conditions required for experimental realization. To delineate the relevant regime, we first consider possible mechanisms that violate the two assumptions underlying the theory, the incompressible limit and the quasi-one-dimensional DW. To maintain the incompressible limit, the emergence of vortices, whose core hosts vanishing Cooper pair density, should be suppressed. Therefore, $H_{\mathrm{max}}$ [Eq.~\ref{eq:max_effi_H}] must be smaller than the first critical field $H_{c1}$, above which vortices start to proliferate. For example, if the hard-axis anisotropy is dominated by demagnetization, the maximum $\eta$ requires the magnetization to satisfy 
\begin{equation}
    M<\frac{\pi H_{c1}}{\sqrt{2}\mu_0},\label{eq:sc_constraint_magnetization}
\end{equation} 
provided that the hard-axis shape anisotropy in thin film geometry is given by $\bar{K}_1=\mu_0M^2$~\cite{rohart2013skyrmion,blundell2001magnetism}. Here $\mu_0$ denotes the free space permeability. Numerically, $H_{c1}>6\,\mathrm{Oe}$ for $M=1\,\mathrm{kA/m}$. In addition, invoking Silsbee's criterion~\cite{tinkham2004introduction}, $H_{\mathrm{c1}}$ should also be sufficiently large to withstand the self-induced magnetic fields generated by the supercurrent. In thin films, the supercurrent density required to destabilize the superconductivity via this mechanism is given by $H_{\mathrm{c1}}/\mu_0\lambda_p$ with the penetration depth $\lambda_p$~\cite{talantsev2015universal}. Since the typical order of $\abs{j_{c}^{\pm}}$ is determined by $\lambda K_1/2\mathcal{P}$~[see Eqs.~(\ref{eq:jp}) and (\ref{eq:jm})], we find the relation:
\begin{equation}
    \bar{K}_1<\frac{\hbar H_{\mathrm{c1}}}{\mu_0\lambda_p\bar{\lambda} e }.
\end{equation}
For $M=1\,\mathrm{kA/m}$ and $\bar{K}_1=\mu_0 M^2$, this relation yields $H_{\mathrm{c1}}>0.12\,{\mathrm{Oe}}$ where $\bar{\lambda}=\sqrt{\bar{A}/\bar{K}}=5\,\mathrm{nm}$ and $\lambda_p=1\,\mathrm{\mu m}$ (Hereafter, the same parameters are used in all numerical calculations).

Secondly, the magnetic fields in SCs, regardless of their time-dependence, are screened in the penetration depth $\lambda_p$. Since this screening can induce the nonuniformity of the magnetic fields, thereby violating the quasi-one-dimensional DW assumption, the geometry dependence of magnetic profiles in SCs should be investigated. By following the approach adopted in the Refs.~\cite{tinkham2004introduction,VODOLAZOV2001125}, which evaluates the magnetic field and vector potential profiles in a superconducting slab, we find that the extension is valid under the geometric condition
\begin{equation}
    \lambda_{\bot}=\lambda_p^2/d> w> d.\label{eq:geometry_constraint}
\end{equation} 
Here, $\lambda_{\bot}$ is referred to as the Pearl length~\cite{pearl1964current}. 
This condition also suppresses the distortion of DWs by demagnetization. Another point worth noting is that, in SCs, supercurrents can be induced solely by external magnetic fields, even in the absence of bias currents, known as screening supercurrents. Specifically, when a uniform $H_z$ is applied to a superconducting slab, an inhomogeneous supercurrent flowing along the $x$ direction is generated, which is given by~\cite{VODOLAZOV2001125}
\begin{equation}
    j_{s,x}=-\frac{A_x}{\mu_0\lambda_p^2}=\frac{yH_z}{\mu_0\lambda_p^2},\label{eq:hzsupercurrent}
\end{equation}
according to the London equation. This inhomogeneous supercurrent density can induce an asymmetry in the DW, such as DW tilting, a phenomenon observed in fast-moving DWs in chiral FMs~\cite{Ryu_2012,PhysRevB.90.184427}. To clarify how the inhomogeneous supercurrent density induces the asymmetry, let us focus on the strip edges. The local mismatch in $X$ between neighboring lattice sites at the edge generates additional effective exchange torques, which lead to a small deviation of $\Phi$ of the DW at the edge, hereafter denoted by $\delta\Phi$. The induced $\delta\Phi$ subsequently produces restoring torques arising from the hard-axis anisotropy and the applied magnetic fields, which counterbalance the STT. To gain a rough estimate of the supercurrent-density gradient at which the rotation and tilting become nonnegligible, we analyze the local torque balance at the DW center at the edge by expanding Eq.~(\ref{eq:spinllg}) to first order in $\delta\Phi$. In the limit $\delta\Phi< 1$, this analysis leads to the relation:
\begin{equation}
    \Delta j_{s,x}< \frac{2e\bar{\lambda}(\bar{K}_1-MH_z)}{\hbar},
\end{equation}
where $\Delta j_{s,x}=c\partial_y j_{s,x}$ denotes the difference in supercurrent density between two layers separated by the lattice constant $c$. By substituting Eq.~(\ref{eq:hzsupercurrent}) into the above equation, we obtain 
\begin{equation}
    H_z< \frac{\bar{K}_1}{M}\left(1+\frac{c\hbar}{2\mu_0 M e \bar{\lambda}\lambda_p^2}\right)^{-1}.
\end{equation}
This equation yields $H_z<12\,\mathrm{Oe}$ for $c=1\,\mathrm{nm}$, where $H_{\mathrm{max}}$ is given by $5.7\,\mathrm{Oe}$. 

Lastly, in addition to the phase slip induced by the DW precession, dissipation can also arise from other mechanisms, such as thermally activated phase slips (TAPS) and quantum phase slips (QPS)~\cite{PhysRevB.78.144502,PhysRevLett.87.217003}. To avoid these additional effects, experiments may require cryogenic temperatures far below the critical temperature $T_c$ and macroscopic system size satisfying the constraints in Eq.~(\ref{eq:geometry_constraint}). Furthermore, thermal fluctuations can have a nontrivial impact on the DW dynamics, especially around $j_c^{\pm}$, as demonstrated in the studies of the phase particle dynamics in Josephson junctions~\cite{tinkham2004introduction}. From the analogy with Josephson junctions, the condition $kT<\hbar I_c/2e$ provides a useful estimate of the characteristic temperature, below which the thermal effects become negligible, given by
\begin{equation}
    T<\frac{j_c w d \hbar}{2ek}\simeq\frac{\bar{\lambda}\bar{K}_1w d}{2k}
\end{equation} 
with the temperature $T$ and Boltzmann constant $k$. This relation results in $T<14\,\mathrm{mK}$, implying the need to increase the cross-sectional area.

\section{Conclusion}
\label{sec:conclusion}

We proposed a route to realize the SDE in FMSCs by exploiting the nonreciprocal dynamics of magnetic textures. In this work, the route was based on the nonreciprocal WB of a magnetic DW, where the DW precession induces phase slips in FMSCs. As a first step, we established theoretical frameworks to describe the nonreciprocal WB in FMs for both pinned and unpinned cases. For the unpinned case, we derived analytical expressions for quantities relevant to the diode effect, such as the critical current densities and the diode efficiency $\eta$, and found that the maximum $\eta$ reaches $1/3$ when the external dc magnetic fields are applied simultaneously along the $y$ and $z$ directions with the same intensity $H_c/2$~[Eq.~(\ref{eq:max_effi_H})]. We showed that the nonreciprocal WB can give rise to diode-like I-V characteristics through the adiabatic SMF. The theory was verified by comparison with micromagnetic simulations, which also provided physical insights into the DW rotation mechanisms in both pinned and unpinned cases. Finally, to extend the analysis from FMs to FMSCs, we demonstrated the equivalence between the magnetization dynamics in the two systems under the incompressible limit, and identified the criteria required for this extension, including the conditions for the incompressible limit and the rigidity of the DW. 

\begin{figure}[!t]
    \centering
    \includegraphics[width=0.9\linewidth]{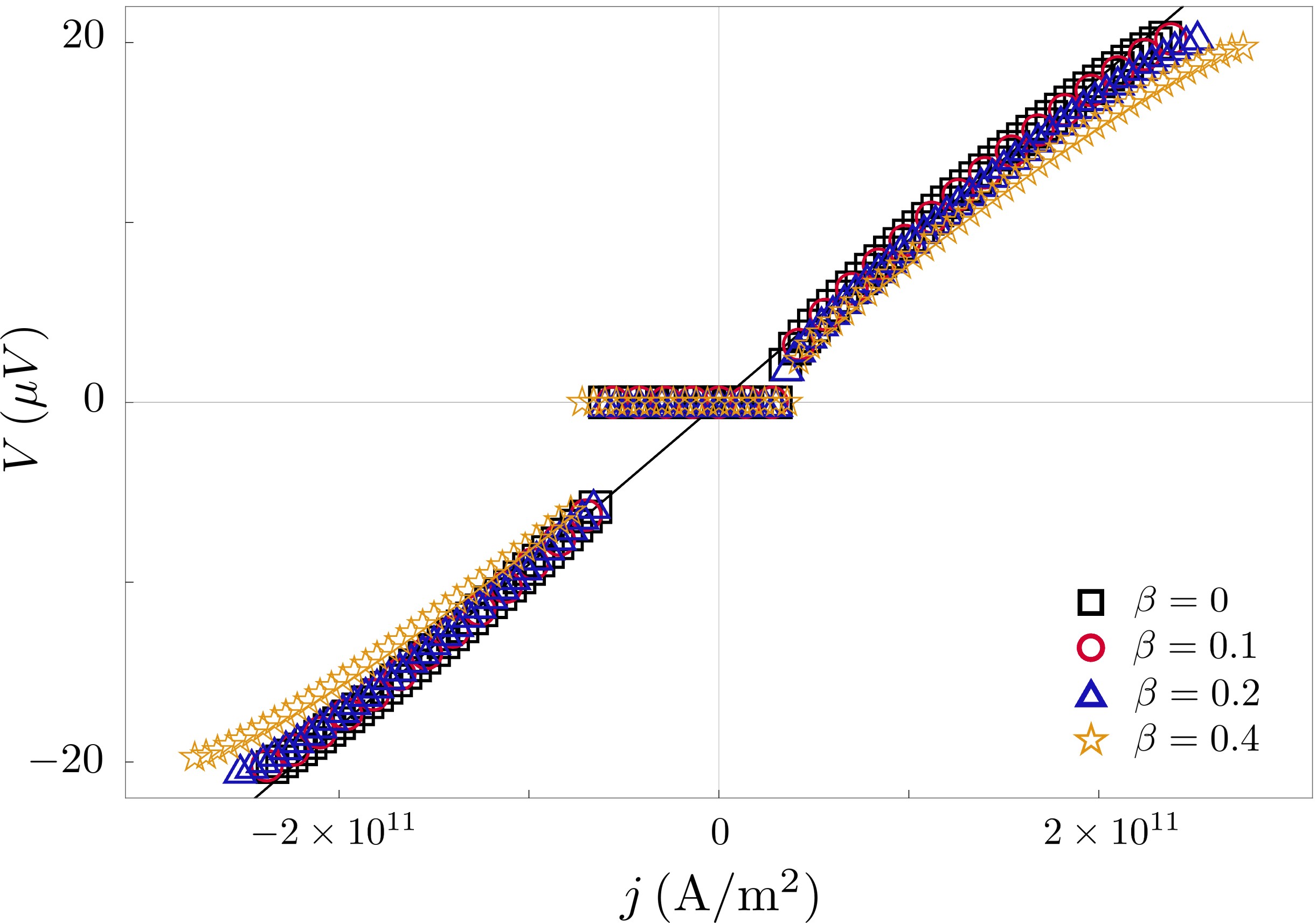} 
    \caption{I-V curves of a pinned DW for various $\beta$ obtained from simulations with $H_y=H_z=30\,\mathrm{Oe}$ and $\alpha=0.2$. The black solid line represents $\langle V\rangle=-2\mathcal{P}\langle\dot{\Phi}\rangle$ [Eqs.~(\ref{eq:smf}) and (\ref{eq:avgs})].}
    \label{fig:nonadiabatic}
\end{figure}

From a broader perspective, this work establishes a topological defect based method to emulate a Josephson junction without heterostructures, providing a single-material platform for realizing Josephson-junction-like effects within superconducting spintronics. This framework also enables the adaptation of concepts established in conventional spintronics. One direction for future research is to investigate the role of the Dzyaloshinskii-Moriya interaction (DMI)~\cite{dzyaloshinsky1958thermodynamic,moriya1960anisotropic}, which arises in the materials lacking IS or at interfaces. The DMI breaking the IS may play a role analogous to that of $H_z$ in the present system. Therefore, in noncentrosymmetric FMSCs such as UIr~\cite{akazawa2004pressure}, the condition for maximizing the diode efficiency $\eta$ may deviate from $H_y=H_z$.

\begin{acknowledgments}

We thank Yaroslav Tserkovnyak, Jae-Keun Kim, and Albert Min Gyu Park for fruitful discussions. S.B.C. thanks the Anthony J. Leggett Institute for Condensed Matter Theory and the Department of Physics of University of Illinois Urbana-Champaign for the hospitality during the sabbatical visit. D.H.H and S.K.K. were supported by the Brain Pool Plus Program through the National Research Foundation of Korea funded by the Ministry of Science and ICT (2020H1D3A2A03099291), National Research Foundation of Korea(NRF) grant funded by the Korea government(MSIT) (RS-2026-25470048), 
Basic Science Research Program through the National Research Foundation of Korea (NRF) funded by the Ministry of Education (2019R1A6A1A10073887), and Creation of the quantum information science R\&D ecosystem(based on human resources)  through the National Research Foundation of Korea(NRF) funded by the Korean government (Ministry of Science and ICT(MSIT)) (RS-2023-00256050). S.B.C. was supported by the National Research Foundation of Korea (NRF) grants funded by the Korea government (MSIT) (NRF-2023R1A2C1006144 and NRF-2018R1A6A1A06024977) and the 2025 sabbatical year research grant of the University of Seoul.

\end{acknowledgments}

\begin{figure}[!t]
    \centering
    \includegraphics[width=0.9\linewidth]{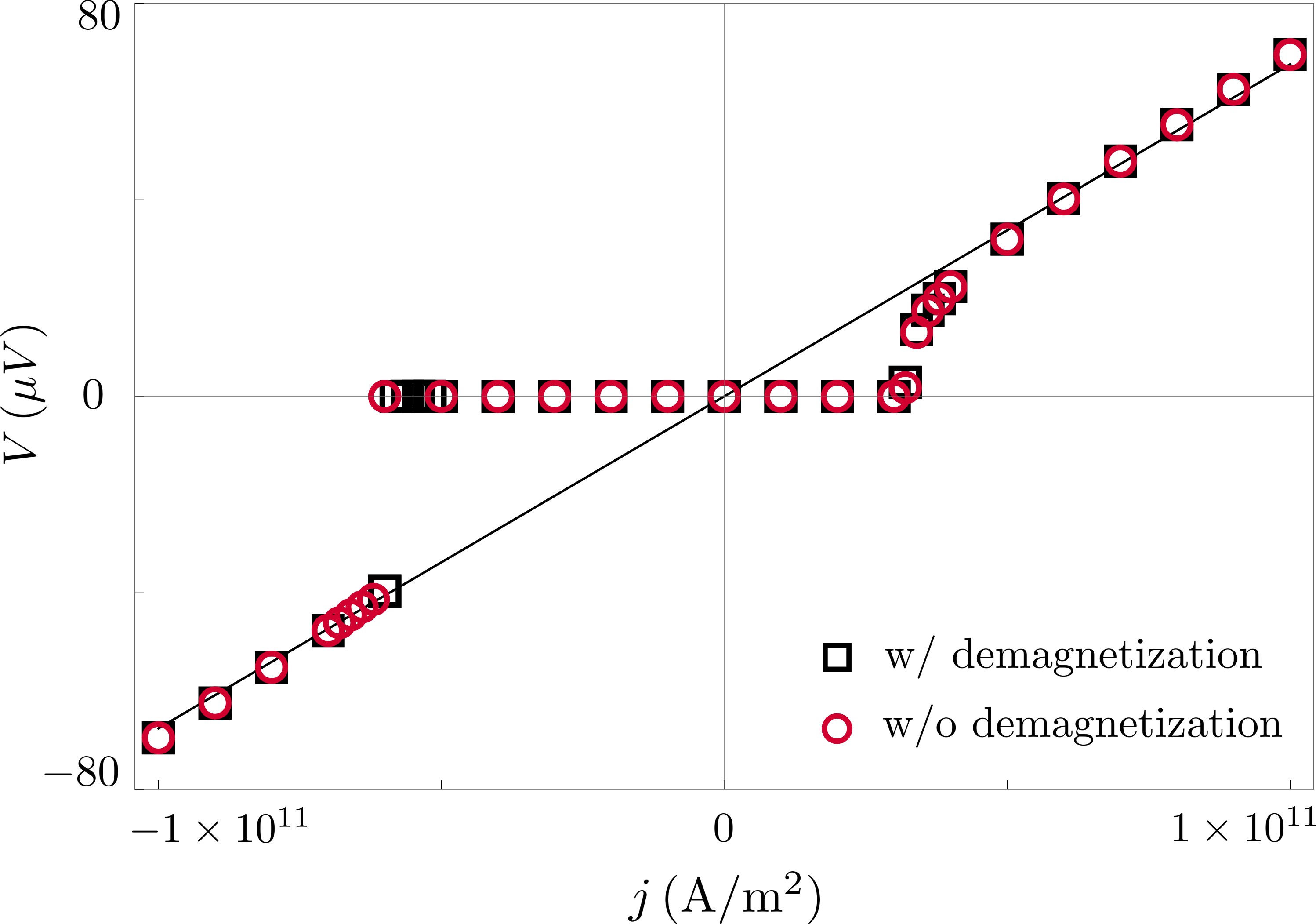} 
    \caption{I-V curves showing the effect of demagnetization on the precessional dynamics of a DW pinned at a notch. Simulations are performed with $H_y=H_z=226\,\mathrm{Oe}$ and $M=40\,\mathrm{kA/m}$, while all other material parameters are the same as those used in the main text. The square symbols represent the case including the demagnetization effect, whereas the circular symbols correspond to the case neglecting demagnetization but including local hard-axis anisotropy. The black solid line represents $\langle V\rangle=-2\mathcal{P}\langle\dot{\Phi}\rangle$ [Eqs.~(\ref{eq:smf}) and (\ref{eq:avgs})].}
    \label{fig:dmag}
\end{figure}

\appendix
\section{Nonadiabatic STT}
\label{app:nonadiabatic}

The nonadiabatic STT, parameterized by $\beta$ in Eq.~(\ref{eq:llg}), reflects the misalignment between the local magnetization and the spin of conducting electrons in realistic systems. In the DW dynamics without pinning, this torque can compensate for the damping effects, thereby suppressing the DW rotation. Especially, for $\alpha=\beta$, there exists a steady-state solution, obtained by a Galilean transformation of the static profile. In this case, the DW undergoes purely translational motion without the precessional dynamics. As $\beta$ deviates from $\alpha$, the DW starts to precess with the handedness determined by the ratio $\alpha/\beta$. Specifically, the nonadiabatic STT modifies Eq.~(\ref{eq:avg_w/o_pinning}) to 
\begin{equation}
    \langle{\dot{\Phi}}\rangle_0=\frac{\mathcal{P}j(\alpha-\beta)}{\mathcal{J}\lambda(1+\alpha^2)}.
\end{equation}
However, the presence of pinning overturns this picture. Equation~(\ref{eq:avgs}) is independent of $\beta$, and the handedness therefore remains unchanged, consistent with the simulation results in Fig.~\ref{fig:nonadiabatic}. This difference stems from the distinct rotation mechanism, discussed in the main text, and the nonadiabatic STT effect on the DW precession is canceled by the enhanced time-averaged displacement from the center. Lastly, the slight flattening of the slope with increasing $\beta$ may arise from the reduced rigidity of the DW due to the excitation of internal modes, such as vortex nucleation within the DW~\cite{nakatani2003faster,NAKATANI2005750,A.Thiaville_2005}.

\section{Demagnetization}
\label{app:dmag}

In the main text, we consider only a local hard-axis anisotropy and neglect the long-range dipolar interactions. To examine the nonlocal effect of the demagnetization, we compare two simulation setups: one without the demagnetization effect but with $K_1=\mu_0 M^2$, and the other without $K_1$ but including the demagnetization effect, since the effective hard-axis anisotropy is given by $K_{\mathrm{eff}}=K_1+\mu_0 M^2$ in a thin film geometry~\cite{blundell2001magnetism}. Figure \ref{fig:dmag} shows that the difference in $j_c^{\pm}$ between the two cases is marginal, and the slope follows the same trend as in the main text. The condition for $\eta_{\mathrm{max}}$ is given by $H_y=H_z=226\,\mathrm{Oe}$ for $M=40\,\mathrm{kA/m}$ [see Eq.~(\ref{eq:max_effi_H})], yielding $\abs{\eta}=0.33\,(0.31)$ without (with) demagnetization. The small deviation may arise from demagnetization effects combined with the notch geometry.

\bibliography{diode_fmsc.bbl}

\begin{thebibliography}{67}
\bibitem [{0_f()}]{0_fmsc_footnote}%
  \BibitemOpen
  \href@noop {} {}\bibinfo {note} {The EOM of $\Phi$ does not contain an inertial term proportional to $\ddot{\Phi}$}\BibitemShut {NoStop}%
\bibitem [{2_f()}]{2_fmsc_footnote}%
  \BibitemOpen
  \href@noop {} {}\bibinfo {note} {For the present N\'{e}el-type DW, the effective potential landscape for $\Phi$, viewed in one of its ground states, becomes dependent on the handedness of the DW precession through $H_z$, which breaks $\mathcal{M}_y:\Phi\rightarrow-\Phi$. Furthermore, if we assume the easy-axis anisotropy $K_1<0$, which changes the ground state from the N\'{e}el-type DW to the Bloch-type DW, the roles of $H_y$ and $H_z$ are interchanged and $\mathcal{M}_z:\Phi\rightarrow\pi-\Phi$ becomes the mirror symmetry relevant to nonreciprocity.}\BibitemShut {Stop}%
\bibitem [{1_f()}]{1_fmsc_footnote}%
  \BibitemOpen
  \href@noop {} {}\bibinfo {note} {Within the notch, as discussed in the main text, the DW position acquires a finite time-averaged displacement $\expval{X}$ and a nonzero but small standard deviation, which justifies the harmonic approximation in Eq.~(\ref{eq:pinningpotential}). By contrast, $\dot{X}$, the time derivative of $X$, and other oscillatory quantities, such as $\sin\Phi$, are assumed to vanish upon long-time averaging.}\BibitemShut {Stop}%
\end{thebibliography}

\end{document}